\documentclass[10pt,letter]{article}

\usepackage{fullpage}

\newcommand{\size}[1] {\vert#1\vert}
\usepackage{amssymb, amsmath,shortvrb,psfrag,graphicx}
\usepackage{subfigure,epsfig}

\begin{document}
\title{Diffusive Load Balancing of Loosely-Synchronous Parallel Programs over Peer-to-Peer Networks}
\author{
Scott Douglas and Aaron Harwood\footnote{Author for correspondence.}\\
Department of Computer Science and Software Engineering\\
University of Melbourne\\
Victoria, 3010. AUSTRALIA \\
{\tt \{scdougl,aharwood\}@cs.mu.oz.au}
}

\maketitle
\begin{abstract}

The use of under-utilized Internet resources is widely recognized
as a viable form of high performance computing. Sustained processing
power of roughly $40T$ FLOPS using 4 million volunteered Internet hosts
has been reported for embarrassingly parallel problems. At the same
time, peer-to-peer (P2P) file sharing networks, with more than 50 million
participants, have demonstrated the capacity for scale in distributed
systems. This paper contributes a study of load balancing techniques
for a general class of loosely-synchronous parallel algorithms when
executed over a P2P network. We show that decentralized,
diffusive load balancing can be effective at balancing
load and is facilitated by the dynamic properties of P2P.
While a moderate degree of dynamicity can benefit load balancing,
significant dynamicity hinders the parallel program performance
due to the need for increased load migration. To the best of our
knowledge this study provides new insight into the performance of
loosely-synchronous parallel programs over the Internet.

\end{abstract}

\noindent {\bf keywords:} peer-to-peer computing, load balancing, loosely-synchronous

\section{Introduction}

As the number and performance of Internet hosts continues to increase,
so does the number of under-utilized resources. These under-utilized
resources are widely recognized as potential processing nodes for
high performance computing (HPC) projects;
\emph{Seti@Home}\footnote{http://www.setiathome.ssl.berkely.edu} reports
gaining $40T$ FLOPS of average processing power using about 4 million
hosts. This is a remarkable use of idle processing power that can't
be understated albeit that the problem is embarrassingly parallel.
This paper addresses the execution of \emph{loosely-synchronous} parallel
programs over a similar number of resources, which poses significant
additional burdens on the distributed system.

There is now a widely proliferating Grid methodology~\cite{foster99grid} that
is being used repeatedly to link HPC centers and other distributed
resources together. The methodology is hierarchical, consisting of
Grid controllers that manage a homogeneous pool of hosts, e.g.
Globus~\cite{foster97globus}, XGrid\footnote{http://www.apple.com/acg/xgrid/}
and Grid brokers that coordinate
jobs over a heterogeneous set of controllers. Brokers compete against or
cooperate with
one another in a computational market~\cite{buyya00high}, selling processing
power to Grid clients. An essential aspect that requires significantly
more research is
the realization of a system architecture that facilitates the
efficient coordination of these resources. The conventional
Grid hierarchy is a \emph{centralized} method of achieving coordination.
While a tree does provide scalability and is suitable for some problems,
e.g. Domain Name Service is quite successful, we believe that a hierarchical
structure will not allow the spontaneous growth of massively parallel
programming that is suitable for all kinds of parallel programs. 

We propose the continued development of \emph{Peer-to-Peer} (P2P)
networks to construct a massively parallel programming infrastructure.
P2P networks provide a completely \emph{decentralized} approach
that emancipates the system from a hierarchy without compromising
scalability. Basically, a host in the Internet runs a peer process
that uses a P2P protocol to connect to a number of existing peers.
A P2P protocol provides efficient data storage and retrieval over
all peers in the network without significant impact from the continual
connecting and disconnecting of peers. By distributing the
coordination overhead among all participating hosts, P2P networks
allow greater scalability and robustness, simply stated: the failure
of a any given host is no more likely to disrupt the system than
the failure of any other host.

In particular, we consider the coordination of resources for the
purpose of maximizing the efficiency of loosely-synchronous parallel
programs. Fundamental to this is the load balancer which should
support the favorable characteristics of the P2P network on which
it runs, namely decentralization. In addition, since P2P networks
consist of hosts that participate for limited time intervals, it
must be able to adapt to the changing network; in other words be
dynamic.

While distributed computing on (semi) P2P networks has been shown to work for
some parametric and data parallel problems such as
\emph{Distributed.net}\footnote{http://www.distributed.net} and Seti@Home in
which load balancing can be accomplished by work pooling, loosely synchronous
problems have the additional challenge of accounting for inter-task
communication. For example, in the simulation of fluid
dynamics~\cite{fox94parallel} each task requires regular synchronization with a
subset of other tasks.  Consequently the overall progress of the application is
restricted to the rate of the slowest task, and therefore the quality of the
load distribution has a significant impact on performance. Furthermore, since
communication latency is affected by the distance between coupled tasks, the
load balancer should be able to account for task locality.

P2P networks have been extensively used for file sharing over the
Internet, for example \emph{Gnutella}, \emph{KaZaa}, \emph{Morpheous}
and \emph{Freenet}. Recent study of P2P architectures such as
Chord~\cite{chord}, Tapestry~\cite{zhao01tapestry} and
Past~\cite{druschel01past} provide structures that enable more
efficient document searching. The flexibility of P2P enables the
formation of structures that are also beneficial to parallel and
distributed computing, namely the discovery of and coordination
of computational resources.

While some work considers execution of load balancing over unreliable
networks~\cite{aiello93approximate} and the execution of parallel
applications, in particular with MPI~\cite{aselikhov20029-th}, the
use of P2P for loosely synchronous programs is not well studied.
This paper investigates the issues involved with the execution of
loosely synchronous programs over unreliable P2P networks, in
particular, we focus on the applicability of \emph{Diffusive load
balancers}.

Diffusive load balancers achieve a global balance by continuously
arranging load within a sub set (or \emph{domain}) of the network.
We consider their use for P2P networks because they are: decentralized,
using only locally available information; dynamic, since static
techniques can not accommodate the dynamic behavior of P2P networks;
applicable to any network; and they are simple to implement.

The remainder of the introduction gives a background on previous
approaches to load balancing. Section \ref{sim} describes the model,
strategies and evaluation metrics used in the simulation, the results
of which are discussed in Section \ref{res}.

\section{Decentralized load balancing techniques}

We model loosely synchronous applications as a graph called the 
\emph{guest} graph \(G\). Each node in \(G\) represents a job and an edge 
exist between two nodes if synchronization is required between them. 
The network is also modeled as a graph, called the \emph{host} graph \(H\). 
Nodes and edges in \(H\) represent hosts and communication channels between them 
respectively.

Load balancing is the process of allocating or \emph{mapping}
each host in \(H\) an amount of work that maximizes the
overall application performance. Typically this involves the collection of
system information, calculation of the optimal arrangement, and the
\emph{migration} of work units or \emph{jobs} between hosts. We
discuss here only load balancing strategies that are relevant to
distributed applications, a broader introduction can be found 
in~\cite{survey} and~\cite{casavant88taxonomy}.

Throughout this paper we refer to \emph{quality} and \emph{stability}.
We use the term quality to indicate the effectiveness of a mapping
with regard to the progress of a parallel program, a good quality
mapping leads to a faster progression of the program. Quality is a
function of processor load and communication delays.  Stability is
the property of a load balancing which indicates whether it reaches
a point where no further migration is performed.

\subsection{Random load balancing}

Two common methods for random load balancing are random pushing and
random pulling.  For random pushing, an over-loaded host  migrates load
from itself to another host chosen at random. For random pulling, an
under-loaded host migrates load from another host chosen at random.
Both methods have been shown to be effective in~\cite{sanders94detailed}.
Commonly, a host is deemed either under-loaded or over-loaded depending on
whether its' load is respectively greater than or less than, a
threshold.  In this way the quality of the balance is sensitive to
the chosen threshold. An adaptation to this method is discussed 
in~\cite{adler95parallel} where a collection of hosts are polled
in order to choose the most appropriate migration destination.

The simplest randomized load balancers do not require any system
wide information, avoiding the overhead associated with collecting
it. Consequently, the method scales well and is thus applicable to
P2P networks. In addition, since exiting hosts (those leaving the
P2P network) perform pushing, the method suits dynamic networks.
However, the approach can suffer from excessive communication when
the system is under-loaded.  Random pulling suffers when there are
not many hosts with load since a randomly chosen host will likely
not have any load to share.  Random pushing suffers when there are
many hosts with load since a randomly chosen host will likely already
have sufficient load.  Furthermore, the distribution of load occurs
without regard to the locality of the application, in other words,
closely coupled jobs may be placed on distant hosts thereby causing
increased latency.

\subsection{Clustering methods}

Assigning an administrative host to a subset of the system allows
each cluster to be balanced by centralized methods. Administrative
hosts negotiate load migration between clusters thus globally
balancing the load. Scalability is achieved by using scalable
protocols for inter-cluster balancing.

This approach suits systems that have a hierarchical structure such
as grids~\cite{vannieuwpoort01efficient} and wide area networks of
work stations~\cite{becker95dynamic}. Grid architectures for example,
often have different communication capabilities between hierarchies
making inter-cluster migration unfavorable.  Several ``cluster-aware''
approaches based on random pulling and pushing that take this into
account are proposed in~\cite{vannieuwpoort01efficient}.

Many favorable characteristic of P2P such as scalability and
anonymity, stem from the absence of hierarchy. Consequently, while
clustering strategies are sensitive to the locality of jobs,
hierarchical load balancing approaches should be avoided.

\subsection{Diffusive methods} \label{diffuse}

Diffusive load balancing, first proposed by~\cite{cybenko} 
and~\cite{boillat}, allows hosts to be members of several overlapping
domains. In this approach the intersecting host of two domains
perform the task of inter-domain balancing. In other words, by
locally balancing overlapping domains a global balanced is achieved.

Corradi, Leonardi and Zambonelli give a useful definition of diffusive load
balancers in~\cite{corradi99diffusive}. A load balancing strategy can be said
to be diffusive when: (\emph{i}) it consists of identically distributed
components acting autonomously and asynchronously, and (\emph{ii}) it balances
the load within its domain as if it were a separate system and based only on
information from this domain, and (\emph{iii}) each local domain partially
overlaps with other domains so that their unification gives full coverage of
the network.

In sender-initiated diffusion (SID)~\cite{cortes99performance} an over-loaded
host migrates load to an under-loaded neighbor. Each domain consists of two
hosts.  For example, if hosts \(i\) and \(j\) have loads of \(w_{i}\) and
\(w_{j}\) respectively, where \(w_{i} > w_{j}\), then \( \frac{w_{i} -
w_{j}}{2}\) load can be migrated to \(j\) to balance the load. Since the
domains overlap, this method performs an optimal mapping, however, it is based
on an unreasonable assumption that the load is continuous.

The model can be adapted to handle discrete load by migrating
\(\lceil \frac{w_{i} - w_{j}}{2} \rceil \) load to \(j\) if \(w_{i}
> w_{j}\), assuming \(1\) to be the unit load.  However, if \(w_{j}
< w_{i} \leq w_{j}+1\), migrating a unit of load to \(w_{j}\) results
in \(w_{i} < w_{j} \leq w_{i}+1\). Since an imbalance remains, the
next balancing step may return the load to \(w_{i}\), we call this
condition \emph{over-migration}. Essentially the load continues to
migrate between the two hosts until it is consumed, consequently
the strategy is unstable.

To avoid this, often strategies only migrate load between two
neighbors \(i\) and \(j\) if \(w_{i} - w_{j} > 1\), however, this
leaves the hosts in an imbalance. While each domain has an imbalance of
at most \(1\) every domain may have such an imbalance resulting in
a global load \emph{gradient}. In general if such a gradient exists,
the maximum imbalance is \(\frac{D_{H}}{D_{d}}\), where \(D_{H}\)
is the diameter of the network and \(D_{d}\) is the diameter of the
domains, i.e. the larger the domain size, the lower the global
gradient and imbalance.

Extensions to this model have been proposed in~\cite{diffusive00elasser}
to accommodate heterogeneous networks and in~\cite{murthuKrishnan98first,
diekmann97engineering} which uses a limited memory of previous
migrations to avoid over-migration. In~\cite{diekmann99efficient}
it is suggested that diffusive methods should be used as a preprocessing
step to avoid needless migrations.

\subsection{Load balancing on P2P}

Distributed hash tables (DHT) such as \emph{Chord}~\cite{chord} do
intrinsically support a kind of load balancing since the hash
function, when applied to a job's unique identifier, will distribute
jobs over the P2P network at random with approximate uniform
distribution.  While the allocation itself requires little overhead,
it can result in a poor quality mapping, since when \(\size{G} =
\size{H}\), there is a high probability that some hosts don't receive
any jobs~\cite{chord}. Stoica, Morris et al. (in more 
detail~\cite{rao03load}) address this by proposing a set of \emph{virtual}
hosts, \(V\), that map to hosts, such that \(|V| > \size{H}\). By
moving virtual hosts between hosts the load is adjusted accordingly,
increasing the quality. This approach cannot take into account the
heterogeneous nature of the network since jobs are mapped to a hash
space and not to appropriate hosts. There is also no real mechanism
to account for locality, although caching has been suggested to
provide locality of data access.

An adaption of this method is discussed by Byers, Considine and
Mitzenmacher~\cite{byers03simple} who avoid the use of virtual
hosts, arguing that it gives an unnecessary increase in the
communication overhead, since hosts need to monitor a greater number
of connections.  Alternatively, they use a set of hash functions
that associates each resource to a number of hosts.  Each host is
polled prior to allocation to choose the most appropriate. This
approach does not consider the locality of communicating jobs either.

An interesting approach to load balancing is discussed by Montrosor,
Meling and Babao\v{g}u in~\cite{messor}. It uses an ant analogy to
find unbalanced pairs of nodes. The authors claim this approach to 
be relevant to P2P networks since it exhibits both; reliability, the 
loss of ants or peers does not greatly the system and 
scalability, since more peers can be easily handled by more ants.


\section{P2P parallel processing model} \label{sim}

This section describes our model for parallel processing on P2P.
Section \ref{H} describes the host network model and its dynamic
behavior. Section \ref{G} discusses the loosely synchronous
application model and section \ref{meth} describes the metrics used
to evaluate the load balancing strategies. We then describe
the three diffusive load balancing policies considered
for detailed analysis in this paper.

\subsection{The P2P network model} \label{H}

Two events affect the topology of the host network (\(H\)). A host
\(v\) may enter the network, \( V(H) = V(H) \cup \{v\} \), or a
host \(u \in H\) may leave the network, \( V(H) = V(H) - \{u\} \).
Typically an entering host, \(u\), immediately connects with a
subset of other hosts which become its neighbors, denoted \(N(u)\).
We assume these connections remain until either \(u\) or \(v \in
N(v)\) leave. The rate at which network events occur is called the
\emph{dynamicity} of the network, measured by \emph{half-life}.

Let \(H_{t}\) denote the host network at time \(t\). A network is said
to have a half-life of $\tau_\epsilon$ for a positive constant $\epsilon \geq 0$ if
there is a constant 
$t_0$ and $t^\prime$ such that for all $t>t_0$ with 
 $\tau_\epsilon=t^\prime-t$ the following holds: 

\[\left|\,\size{ H_{t'} \cap H_{t} } - \frac{ \size{H_{t'} } }{2}\right| \leq \epsilon. \]

In the instance of an unexpected disconnection, all data on the
offending host is lost, disrupting the application performance. We
consider this a distinct problem and assume that upon departure
each host passes all necessary data and processed work to its
neighbors.  Likewise, the model assumes that partitions in the
network do not occur.

P2P networks that connect peers according to some strategy, and do
not use ad-hoc connections, are termed \emph{structured}. Chord is
an example of a structured P2P network which has been shown to have
a distance between hosts of 
of\footnote{Throughout the paper \(\log\) is used to mean \(\log_{2}\)}
 \(O(\log |H|)\). Our model is
intended to represent a structured P2P network, it is therefore
assumed the organization strategy can produce a structure that has
a similar topological properties. However we did not want to choose a
particular protocol, such as Chord, so as to avoid anomalies that may
arise that are artifacts of the protocol. Therefore
we use a simple, favorable and realistic strategy:
when a host enters the network it connects to \(\log |H|\) other
hosts at random. 

In order to isolate the effect of dynamicity the size of the host
network is kept constant.  This is achieved in simulation by forcing
$\epsilon\rightarrow 0$ such that a host can only enter the network
when another host leaves and vice versa. While this is unlikely to
occur in practice, we believe that it does not detract from the generality
of the experiment.  As discussed in~\cite{loh96how}, two topological
characteristics that affect load balancing are the average degree
and average distance. While the topology continues to change for
dynamic networks using the above strategy, the degree and average
distance are bounded since the probability that a node survives $k$
half-lives is $\frac{1}{2^k}$. They form the initial network graph
in each of our experiments.

\subsection{The application model} \label{G}

An application is classed as loosely synchronous if its jobs require regular
communication with other jobs. For example, finite element methods divide a
large space into smaller, connected, parts. Each part needs only to synchronize
with its neighbors. Loosely synchronous applications account for a wide variety
of useful computations that can significantly benefit from parallelization.  We
model a loosely synchronous application by a graph \(G\), called the guest
graph.  Each node in the graph represents a job (for example an element) and an
edge exists between two nodes if they require synchronization.

There are four actions that a guest host may perform: \emph{run},
\emph{synchronize}, \emph{block} and \emph{end}.  A running job
\(A\) continues to run until it reaches a synchronization point,
upon which it begins communicating with its neighbors in the guest
graph. If any of \(A\)'s neighbors are not ready to synchronize,
job \(A\) enters the blocking state in which it remains until it
has synchronized with each of its neighbors. After synchronization
it changes to the running state.  Upon completion of an application,
all jobs simultaneously end.  


The granularity size of a loosely-synchronous application is
determined by the ratio of time spent communicating to time spent running
(large grained applications spend more time running).  Smaller
grained loosely-synchronous applications are less applicable to
distributed networks since they are restricted by communication
latency. However, continued improvements to communication technology
does broaden the class of applications that are suited to distributed
networks.

\subsection{Evaluation metrics} \label{meth}

We define and use two different evaluation metrics: the \emph{standard deviation in load} 
 and \emph{application progress}.

The standard deviation, \(\sigma\), is used to measure the load imbalance. If \(w_{i}(t)\) is the
load of the \(i^{th}\) host at time \(t\) and \(\Bar{W}(t)\) is the average load at time \(t\), then
\(\sigma\) is given by:

\[ \sigma(t) = \sqrt{ \frac{1}{\size{H}-1} \sum_{i=0}^{\size{H}} (w_{i}(t) - \bar{W}(t))^{2} } \]

A standard deviation of zero represents an optimal mapping where
each host has exactly the same load.  The rate of convergence of
the standard deviation effects the application performance, since
the faster the optimal arrangement is reached, the more work is
performed.  In dynamic networks where network events may increase
the standard deviation at indeterminate intervals, if the load
balancer is incapable of repairing the imbalance (reducing the
standard deviation) fast enough, the optimal arrangement may never
be reached.

For diffusive strategies the greatest rate of reduction in the
standard deviation is when the local imbalance is high, as at the
beginning of an application when jobs enter the network on a single
host.  Therefore the rate of change of \(\sigma\) decreases as the
network becomes more balanced, resulting in near\footnote{The optimal
will eventually be reached.} asymptotic behavior.

The progress of loosely synchronous applications is limited to the
slowest job, thus it may be more applicable to measure this progress
as opposed to the fairness of the load mapping. If \( I_{j}(t) \)
is the number of times job \(j\) has entered the synchronizations
state at time \(t\) then the average progress, \(\bar{P}\), is given
by:

\[ \bar{P}(t) = \frac{1}{\size{G}-1} \sum_{j \in G} { I_{j}(t) } \]

Unlike standard deviation the progress at one time is indicative
of the performance of the application until that time. Consequently,
after a set time period the progress of an application using two
different load balancers can be compared to indicate their respective
performance.

The number of migrations, \(M\), made by a load balancing strategy
is important because excessive migration inhibits the application
performance. If \(M_{u}(t)\) is the number of jobs that have been
migrated from host \(u\) at time \(t\) then the number of migrations
is:

\[ M(t) = \sum_{u \in H}{M_{u}(t)} \]


In the following sections we describe the load balancing policies used in this paper.

\subsection{Extended Neighbor}

The Extended Neighbor (EN) strategy is discussed by Corradi,
Leonardi and Zambonelli in~\cite{corradi99diffusive}. It is an
extension of the sender-initiated strategy and is parameterized by
the size of the domain. EN \(x\) represents the EN strategy where
each domain consists of all hosts a distance of \(x\) from the
central host. In other words, if \( \size{Path_{ij}}\) denotes the
length of a path between hosts \(i\) and \(j\) then the domain of
a central host \(i\) is given by:

\[ D_{i} = \{ j \mid j \in V(H), x \geq \size{Path_{ij}} \} \]

If \(i\) is the central host and \(j\) is the least loaded host
then the EN strategy migrates \( \lfloor \frac{L_{i} - L_{j}}{2}
\rfloor \) load \(j\). If the central host is the least loaded host
in the domain then a proportional amount of load is migrated from
the most loaded host in the domain. Since load is migrated only if
doing so reduces the imbalance, a gradient may occur in the domain.

Note that when simulating this algorithm any domain value of \(x\)
may be used, however, in practice the communication overhead involved
with larger values becomes prohibitive.  With values of \(x\) that
are greater than the diameter of the graph the strategy behaves
like a centralized method and because it produces an optimal load
mapping it is useful for purposes of comparison.

\subsection{Diffusion Algorithm Searching Unbalanced Domains}

The DASUD strategy proposed by Cort\'es, Ripoll, Senar and Luque
in~\cite{cortes99performance} differs from the EN strategy in that
the minimum and maximum hosts in the domain are used to move load
regardless of whether the central host is one of them. The DASUD
strategy uses only the direct neighbors of the central host as the
domain. This permits a maximum difference between hosts in the
domain of 1 unit of load, which means that the local gradient is
half as steep as the EN 1 strategy.

The main difference between the DASUD and EN strategies is the
amount of overhead incurred in coordinating the domain since the
DASUD algorithm can effectively achieve the same quality mapping
of the EN 2 strategy with significantly less communication.

\subsection{Probabilistic migration} 

Both the EN and DASUD strategies migrate load only if doing so
reduces the imbalance between the two hosts. Doing this makes them
stable and results in a global gradient limiting the quality of the
mapping. The PM \(x\) (probabilistic migration) strategy allows
load to be migrated even if it doesn't reduce the imbalance according
to the probability \(x\). In other words, load is migrated between
hosts \(i\) and \(j\) when \( \size{w_{i}-w_{j}} \leq 1 \) with a
probability of \(x\). By occasionally over-migrating load the global
gradient can be reduced because it allows highly loaded domains to
diffuse their load to less loaded neighboring domains.

A tradeoff between stability and quality exists since allowing a
gradient reduces the quality of the strategy, while reducing the
gradient results in greater stabilization time. The PM strategy
parameterizes this tradeoff by assigning a probability to the
likelihood of migrating load when it does not reduce the imbalance.

\section{Simulation and analysis} \label{res}

This section details the simulation experiments. Section \ref{pm} explores the effect of various 
PM parameters on the standard deviation and application progress. In Section \ref{migcost} the 
number and effect of migrations is looked at, in particular in regard to the progress. The effect 
of the size of the application in relation the size of the network is discussed in \ref{cover}.
Section \ref{dynamicity} explores the effect dynamic networks have on the three diffusive strategies and
section \ref{select} explores the benefit on progress of job selection. All results are averaged over 50 
trials.

\subsection{Probabilistic Migration} \label{pm}

\begin{figure}[h!]
  \centering
  \subfigure[\label{exp3_2}]{
    \centering
    \includegraphics[width=5cm,angle=-90]{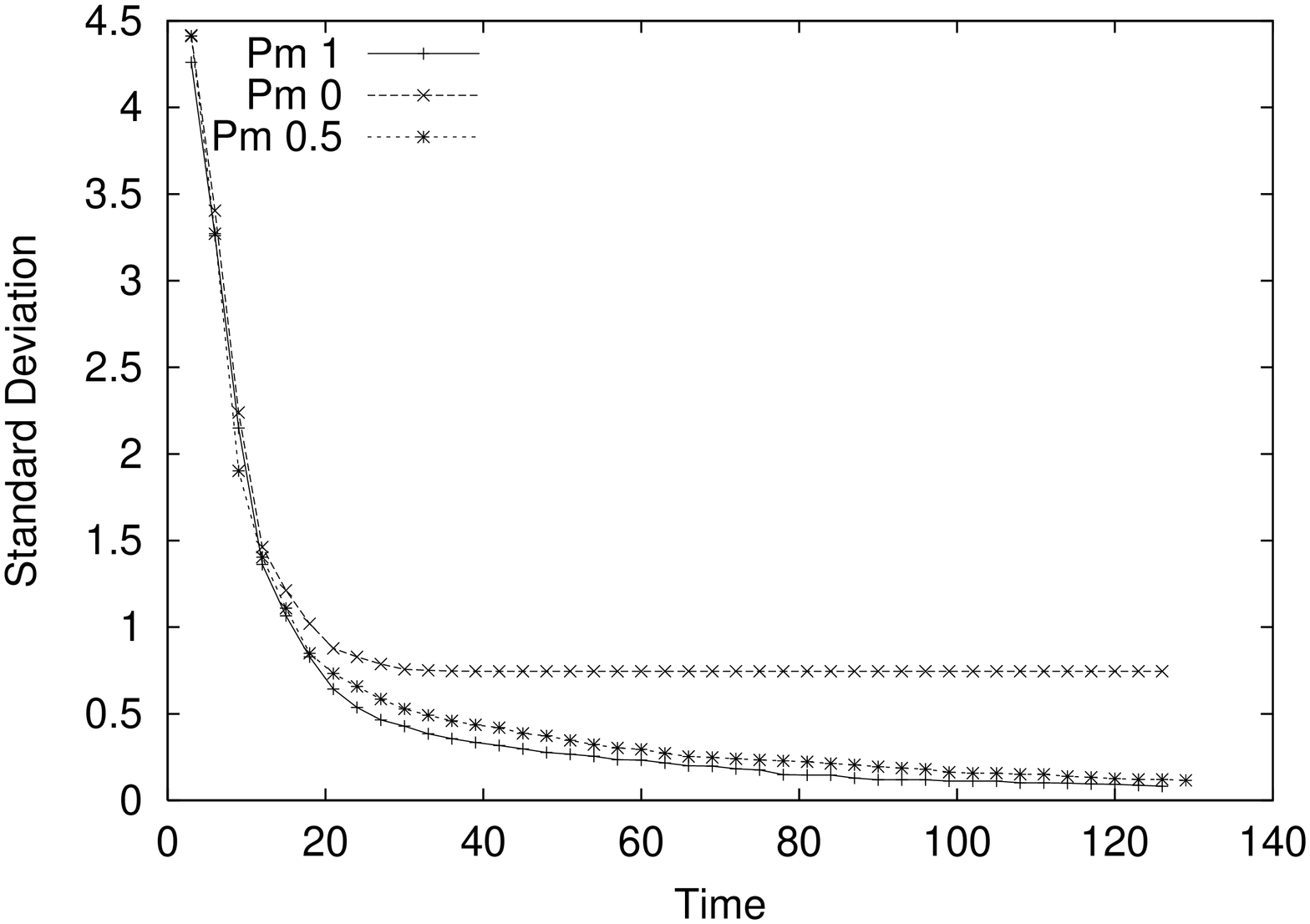}   
  }
  \hspace{10pt}
  \subfigure[\label{exp3_1}]{
    \centering
    \includegraphics[width=5cm,angle=-90]{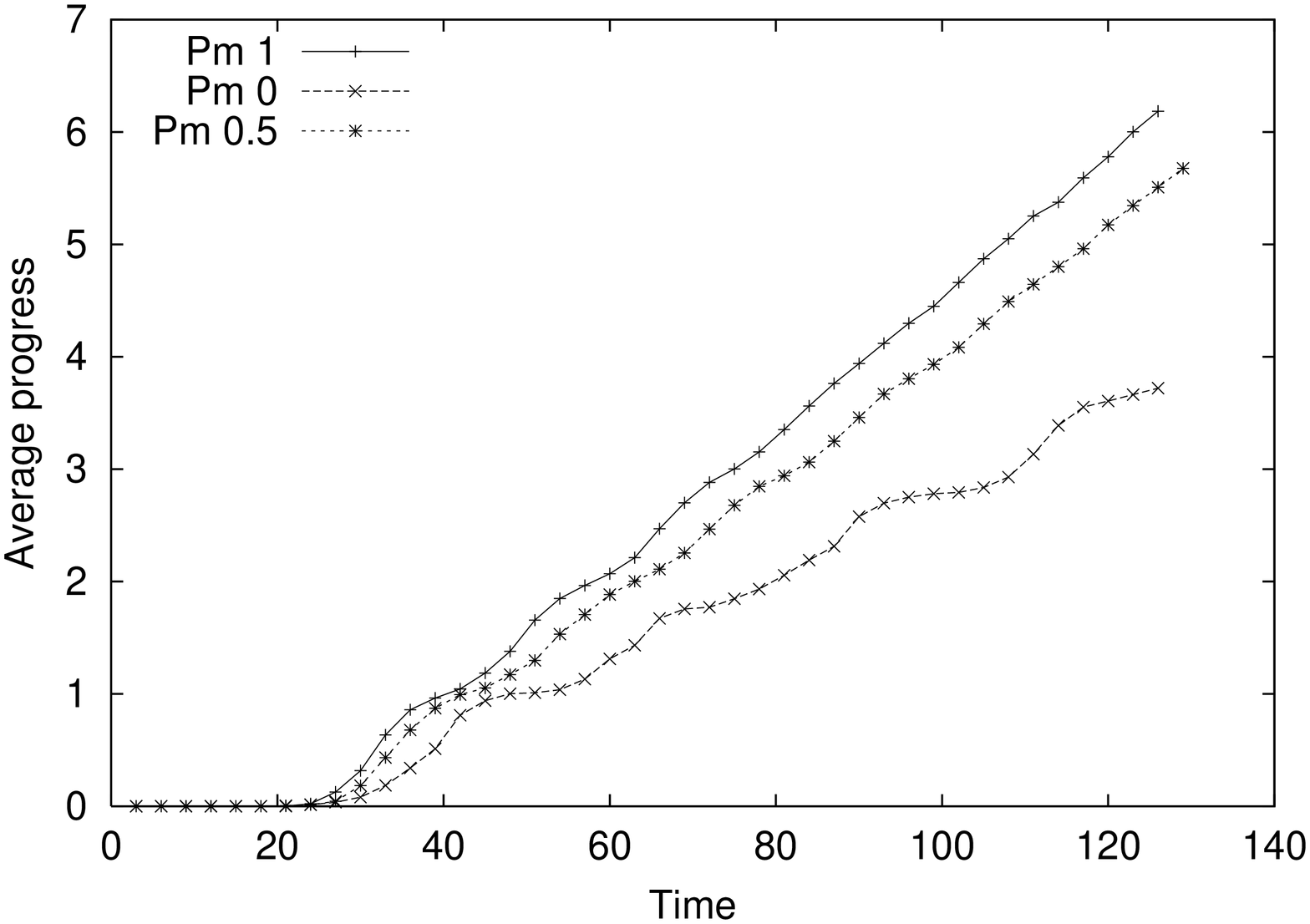}   
  }
  \caption{ A comparison of the PM parameters using the standard deviation (a) and average progress (b).
    \label{exp3}}

\end{figure}

Fig. \ref{exp3} shows the standard deviation and progress for various PM parameters. It uses 
a static graph, with an equal number of jobs to hosts.

The PM 0 strategy migrates jobs from a host only when it does not leave the host with less work 
than its neighbor thus allowing a gradient. It is stable and behaves exactly the same as the EN 1 strategy.
The standard deviation in Fig. \ref{exp3_2} levels off at close to one for the PM 0 strategy, consequently
the progress is significantly less than the optimal (Fig. \ref{exp3_1}). 

The PM 1 strategy always migrates jobs to under-loaded neighbors even if it does not reduce the 
gradient. This effectively shuffles the jobs around the network at random until 
all hosts have an equal amount of work. If the number of jobs and hosts is the same 
(as is in Fig. \ref{exp3}) then the optimal arrangement is eventually found and therefore the strategy 
is stable. Otherwise, the remaining jobs are passed around the network indefinitely and the strategy is 
unstable.

By only over-migrating with a probability of 0.5 the PM 0.5 strategy becomes a hybrid of the previous two. 
This means that in general it won't find the optimum arrangement as quickly as the PM 1 balancer, 
however it does find it eventually and makes approximately half as many migrations when the number of 
jobs and hosts is unequal.

\subsection{Migration cost \label{migcost}}

\begin{figure}[h!]
  \centering
  \includegraphics[width=5cm,angle=-90]{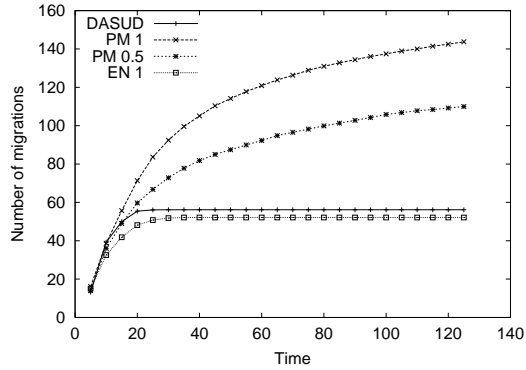}   
  \caption{ The number of migrations for the DASUD, PM 1, PM 0.5 and EN 1 strategies. Using
    a static graph. \label{exp5.2}}
\end{figure}

The disparity in the number of migrations between the load balancing strategies is made 
clear by Fig. \ref{exp5.2}. The number of migrations made by the stable DASUD and EN 1 strategies 
does not increases after time \(t=25\), while, the PM strategies clearly continue 
to make migrations.

\begin{figure}[h!]
  \centering
  \includegraphics[width=6cm,angle=-90]{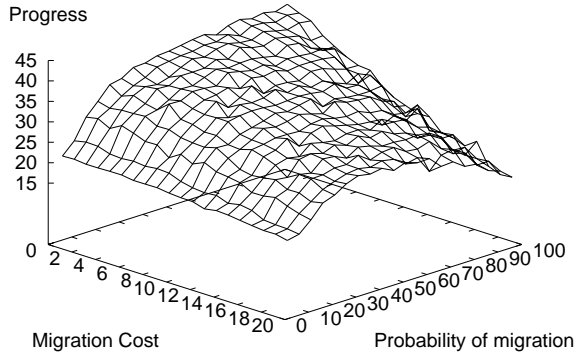}   
  \caption{The number of migrations vs. the probability of migration in relation to the 
    progress at \(t=500\). \label{exp5.3}}
\end{figure}

Migration cost is the time taken for a job to migrate from one machine to another, during which no work 
can be performed on that job. The tradeoff between stability and quality represented by the PM parameter 
provides a means of measuring the relationships between migration cost and progress. 
Fig. \ref{exp5.3} is graph of PM probability vs. migration cost in terms of application progress 
at time \(t = 500\). It shows that migration cost has a greater effect on the progress for high 
probabilities. The most robust performance over a variation in the migration cost occurs at a 
value of approximately 0.35.

\subsection{Application coverage} \label{cover}

Stability is particularly important when the number of jobs does not equally divide into 
the number of hosts which means there are always some edges over which an imbalance exists. 
When a mapping is stable, jobs remain on a host regardless of whether their neighbors are 
under-loaded, otherwise, load may be migrated across these edges even though this cannot remove
the gradient over all edges, consequently the migration does not improve the mapping quality and
thus disrupts the application performance.

\begin{figure}[h!]
  \centering
  \includegraphics[width=5cm,angle=-90]{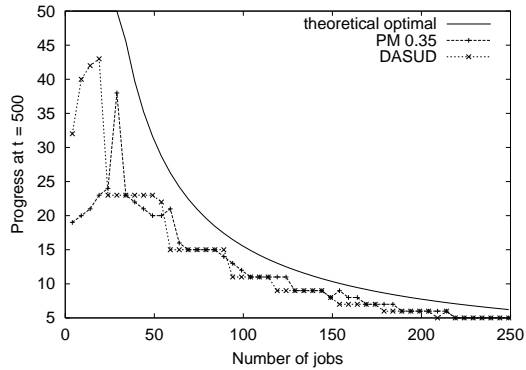}   
  \caption{ A comparison of the DASUD and PM 0.35 strategies in relation to the
    progress at \(t = 500\). With a migration cost of \(5\), \(\size{H}=31\). \label{exp7}}
\end{figure}

The best application performance is achieved when \(\size{G} \leq \size{H}\), using a stable strategy. 
In this case each host gets an equivalent amount of CPU time from each host. When
\(\size{G}+1 = \size{H}\) one host has twice as many jobs and so the 
application progresses at half the average rate as when each host has exactly one job.
This means that as the number of jobs increases the application performance degrades in a 
stepped fashion which can be seen in Fig. \ref{exp7}. 
The line labeled theoretical optimal is the performance expected if load were continuous and 
is defined as $50$ if $\size{G} < \size{H}$ and otherwise as $\frac{\size{G}}{\size{H}}$.
With no migration cost the PM 0.35 strategy has a closer fit to the theoretical optimum than the DASUD 
strategy, however, when a migration cost\footnote{By comparison, on a dedicated host, each 
iteration takes 10 time units.} of \(5\) is used as in Fig. \ref{exp7} it performs better than the DASUD 
strategy only when \(\size{G}\) is close to but not more than \(\size{H}\).

\subsection{Network dynamicity} \label{dynamicity}

A host graph becomes dynamic when hosts enter and exit the network. Such network events alter the load 
distribution and the effect this has on the application is determined by the rate at which the load balancer 
can repair these imbalances. 

\begin{figure}[h!]
  \centering
  \subfigure[\label{exp8_1}]{
    \centering
    \includegraphics[width=5cm,angle=-90]{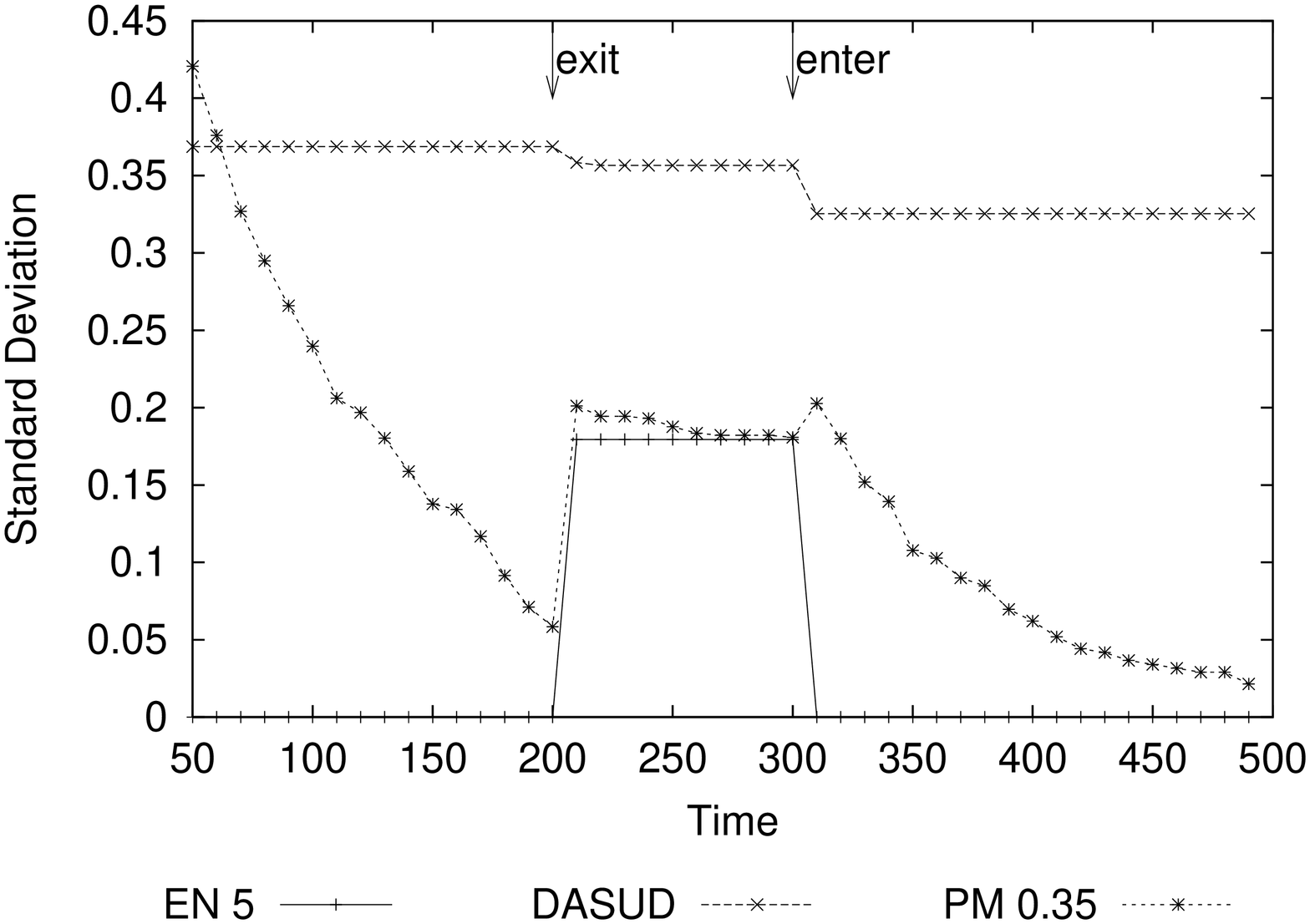}   
  }
  \hspace{10pt}
  \subfigure[\label{exp8_2}]{
    \centering
    \includegraphics[width=5cm,angle=-90]{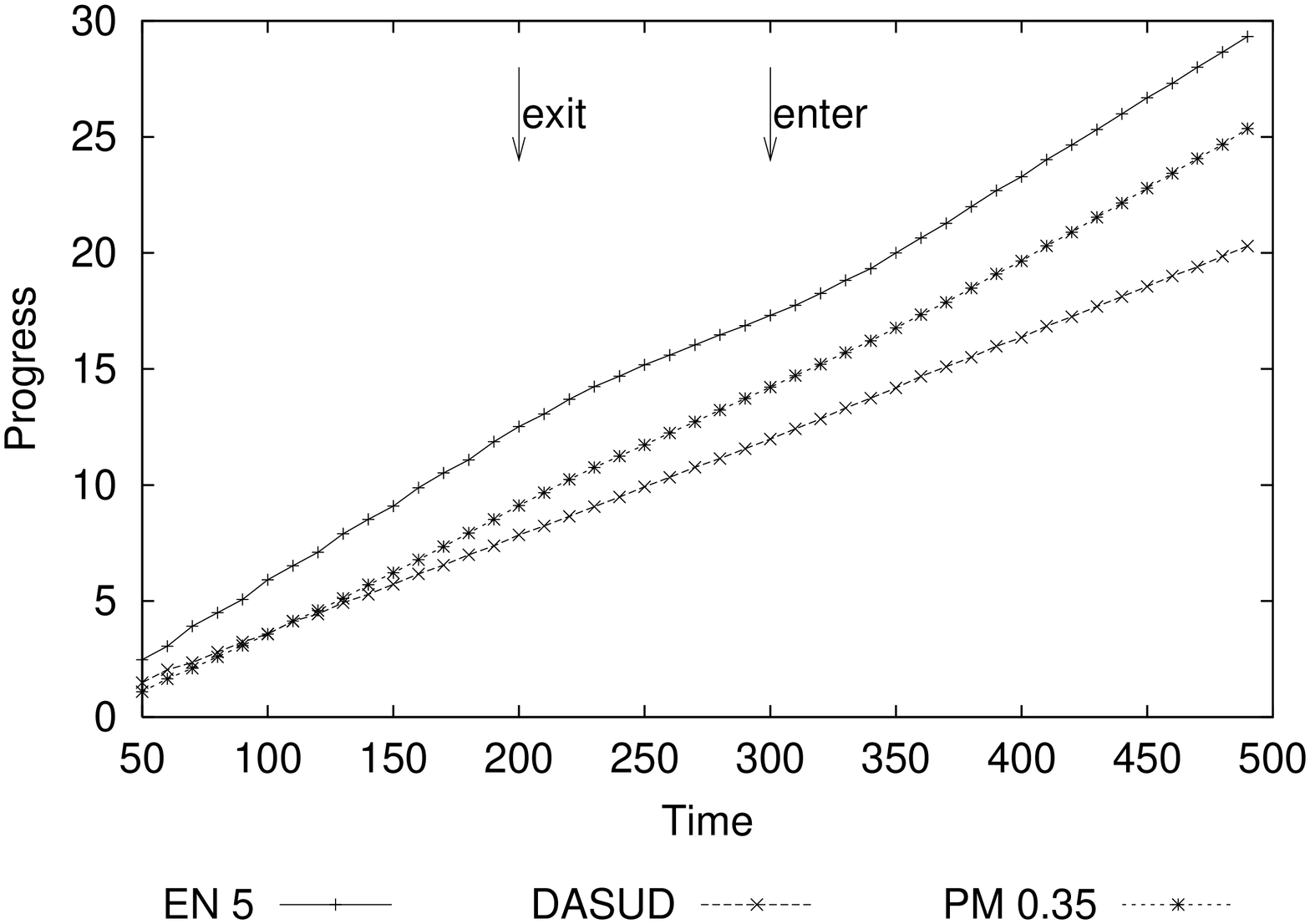}   
  }
  \caption{ The affect of network events on the standard deviation (a) and progress (b)
  using the EN 5, DASUD and PM 0.35 strategies.  \label{exp8}}
\end{figure}

Fig. \ref{exp8} shows the standard deviation (a) and the progress (b) for a 
network in which a host exits at time \(t=200\) and another enters at time \(t=300\). 
For the EN 5 strategy which implements an optimal mapping, an exiting host causes an increase in the 
standard deviation since where before each host had the same number of jobs, now at least one has 
more than the others. The standard deviation is also increased in the PM 0.35 strategy, however, the 
DASUD strategy results in an improved mapping. 

On static networks the DASUD strategy has a local gradient of \(\frac{1}{2}\) but when the load of an 
exiting host is pushed into the network it is altered. If a job is migrated to a domain with such a 
gradient, the DASUD strategy will re-map the local domain, resulting in a reduced gradient and thus a 
reduced standard deviation. In other words, exiting hosts randomize the local load distribution allowing 
the balancer to reduce the gradient. 

An entering host connects to \(log \size{H}\) neighbors at random and creates a new domain. 
The connection of two hosts to the new host reduces the distance between them to at most the
diameter of the domain. When this occurs to two hosts with an imbalance the result is a better 
mapping and hence a reduction in the standard deviation.

While the standard deviation improves with network events for the DASUD strategy little difference 
is made to the progress, as shown by \ref{exp8_2}, since a gradient still exists. 
The progress, when balanced by the PM 0.35 strategy, is effected by the events but less obviously than 
for the optimal EN 5 strategy. When a host exits at least one host has more work than others which 
reduces the progress. When a host enters, the optimal mapping is restored and the progress rate 
returns to its value before the exit event. 

A similarity exists between dynamic networks and the PM strategy since both network events 
and over-migration result in a greater number of migrations but also a better quality mapping. 
In other words, increased instability in the load balancing strategy and increased dynamicity in 
the host network have a similar effect on the applications performance. 

\begin{figure}[h!]
  \centering
  \includegraphics[width=5cm,angle=-90]{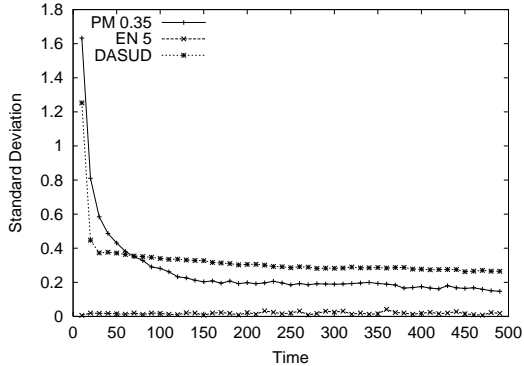}   
  \caption{ The affect of network dynamicity on the standard deviation for
    \(\tau=1000\) using the EN 5, DASUD and PM 0.35
    strategies. \label{exp9}}
\end{figure}

Networks with a low half-life (highly dynamic) suffer from a greater number of migrations. In addition, 
if the balancing strategy is unable to re-map the load between network events fast enough  
longer periods will be spent in an imbalance, resulting in less productive applications. 
The graph in Fig. \ref{exp9} shows the load balancing performance of the EN 5, DASUD and 
PM 1 strategies in terms of standard deviation and progress for a network with \(\tau = 1000\). 
The dynamicity has a positive effect on the DASUD strategy which is evident 
by the consistent reduction in the standard deviation throughout the experiment. The PM 1 
strategy does not reach an optimum balance as it would on a static network indicating that it is 
unable to repair imbalances between network events. This is a consequence of its near asymptotic 
rate of convergence.

\begin{figure}[h!]
  \centering
  \includegraphics[width=5cm,angle=-90]{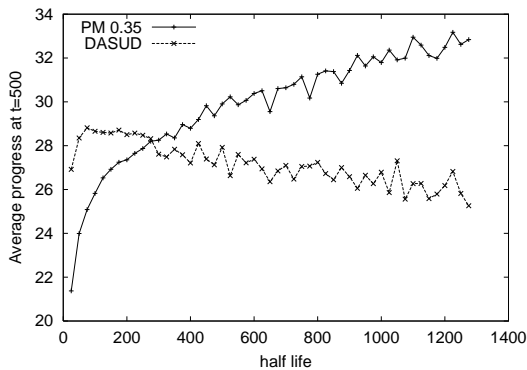}   
  \caption{ The effect of network dynamicity on the averaged application progress at time 
  \(t=500\) for the DASUD and PM 0.35 strategies.\label{exp10}}
\end{figure}

The fact that the DASUD strategy has a lower standard deviation on dynamic networks is a 
result of the randomizing behavior of network events. Entering and exiting hosts re-distribute 
the local load which alters the load gradient, positively effecting the global balance.
However, too much dynamicity impairs the application performance due to a greater time spent 
imbalance and an increase in the number of migrations. Fig. \ref{exp10} shows the effect 
of host dynamicity on the application progress at \(t=500\) for the DASUD and PM 0.35 strategies. The 
DASUD strategy has the best performance when \(\tau\) is approximately \(50\) re-affirming that 
dynamicity can have a positive effect. Networks with more dynamicity than this have a 
detrimental effect implying that this level of dynamicity has an average network interval 
roughly equivalent to the stabilization time of the strategy. The effect of highly dynamic 
networks on the PM 0.35 strategy is more dramatic since its instability incurs more 
migrations. For less dynamic networks however the performance is better on average than the DASUD strategy.

\subsection{Job Selection} \label{select}

In the previous experiments the time required to synchronize jobs
is constant, regardless of the distance between them. This is
unrealistic since latency increases with the distance between hosts.
While it is also dependent on a number of other factors, not least
of which is the network traffic, for the purposes of simulation we
introduced a synchronization latency that is dependent on the number
of hops it makes. In this way the synchronization time of a job is
proportional to the maximum distance between the hosts of neighboring
jobs.  In other words, if \(u \in H\) is the host of job \(i \in
G\) and \(V\) is the set of hosts of each of \(i\)'s neighbors,
then the synchronization time of \(i\) is \( S_{i} = \gamma \max{
\{ \size{PATH_{uv}}\, \mid v \in V\, \}  } \), where \(\gamma\) is a
constant representing the cost per hop.

\begin{figure}[h!]
  \centering
  \includegraphics[width=5cm,angle=-90]{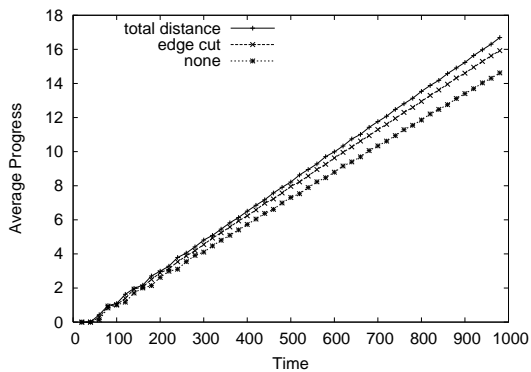}   
  \caption{The effect of job selection on progress using minimum
  total distance, minimum edge cut and none.  With the PM 0.35
  strategy and \(\gamma=3\). \label{exp11}}
\end{figure}

Previously jobs were selected for migration at random. However,
when using the latency model above, by taking into account the
locality of neighboring jobs, an increase in the application progress
can be made, evident by Fig. \ref{exp11}.

Three selection methods are considered: \emph{minimum total distance}, \emph{minimum
edge cut} and \emph{none}. Minimum distance selects the job that will reduce
the maximum total distance when migrated to a neighboring host.
It represents the best possibly reduction in latency for migration
between two hosts, however, in practice the overhead required in
measuring this value could be very large, since it involves measuring
the distance between each neighboring pair of jobs on both the
current host and the proposed host. To avoid this overhead the
minimum edge cut selects the job with the smallest edge cut on the
proposed host.  The edge cut is defined as follows: If \(u,v \in
H\) are the hosts of jobs \(i,j \in G\) respectively then the edge
cut of a job \(i\) is given by \( C_{i} = \size{ \{ j \mid u \neq
v, j \in N(i) \} } \). As expected this approach does not perform
as well as the minimum total distance, however, it does give a
marked improvement over no job selection without excessive overhead.

\section{Conclusion}

This paper contributed a parallel processing model for loosely-synchronous
parallel programs executing over a P2P network. The network was
modeled as a host graph and the program was modeled as a guest
graph. We compared the quality of a number of relevant load balancing
strategies, where good quality leads to a faster rate of progression.

We have shown that for loosely synchronous applications the quality
of the mapping is important, since the average progress of the
application is limited to the rate of the slowest job. Stable
diffusive strategies lead to an undesirable global load gradient
and hence are a poor quality mapping. With unstable strategies,
migrations may continue indefinitely and the relative improvement in
quality is no longer significantly proportional to the number of
migrations. Consequently, the performance of the application
deteriorates as the migration cost increases, especially for
applications that do not cover the network.

The PM strategy parameterizes the tradeoff between quality and
stability. The result of varying PM parameter shows that the most
robust value is around 0.35 indicating that a little over-migration
improves performance.

For stable strategies with a local gradient, network dynamicity can
result is an improvement in the quality. This is a consequence of
exiting hosts temporarily disrupting the local gradient, effectively
randomizing the local load in a similar manner to the PM strategy.
Entering hosts facilitate the load balancing by joining randomly
selected hosts which in turn can reduce the distance between 
host pairs that are in an imbalance.

If the stabilization time of the load balancer is less than the
average period between network events then the mapping will never
stabilize. Therefore, strategies with fast stabilization times
perform better for highly dynamic networks even if the quality is
less than optimal. The PM strategy has better application performance
than the DASUD strategy only for static or slightly dynamic networks
and when the number of jobs divides equally into the number of
hosts.

Diffusive load balancing strategies have been shown to be applicable
to P2P networks, with the dynamic behavior of the network having
a positive effect on the quality of the mapping. For loosely
synchronous applications, however, the inherent gradient involved
with the stable diffusive strategies results in poor application
performance.

We plan to extend our simulations to realistic internets using a
more detailed simulation package and to make use of real P2P protocols. 
This will allow the study of heterogeneous networks, with greater analysis 
on the affect of job communication and host capacity. Since P2P 
networks may consists of non-dedicated hosts this capacity may change 
over time. In other words, not all load is migrateable.

Further theoretical analysis is also needed in order to determine the most
appropriate organizational strategy to form efficient structures and to 
better understand the limit to which network dynamicity has an beneficial 
affect on the quality of diffuse load balancing.

\bibliographystyle{abbrv}
\bibliography{references}  

\end{document}